\documentclass[a4paper,11pt]{article}
\pdfoutput=1 
\usepackage{jcappub} 
\usepackage{array}
\usepackage{multirow}
\usepackage{float} 
\usepackage{xcolor}
\usepackage{natbib}
\usepackage{graphics}
\usepackage{graphicx}
\usepackage{color}
\usepackage{ifthen} 
\usepackage{mathrsfs}
\usepackage{amsmath, amssymb}
\usepackage{tabularx}
\usepackage{subcaption}
\usepackage{titlesec}

\def\ds{{\rm d}s}
\def\da{{\rm d}a}
\title{\boldmath Local patch analysis for testing statistical isotropy of the Planck convergence map}

\author[a,b,1]{Priya Goyal,\note{Corresponding author}}
\author[a]{Pravabati Chingangbam}

\affiliation[a]{Indian Institute of Astrophysics, Koramangala II Block,       
  Bangalore  560 034, India}
\affiliation[b]{Department of Physics, Pondicherry University, R.V. Nagar, Kalapet, 605014, Puducherry, India}

\emailAdd{priya.goyal@iiap.res.in}
\emailAdd{prava@iiap.res.in}
\abstract{The small but measurable effect of weak gravitational lensing on the cosmic microwave background radiation provide information about the large-scale distribution of matter in the universe. 
We use the all sky distribution of matter, as represented by the {\em convergence map} that is inferred from CMB lensing measurement by Planck survey, to test the fundamental assumption of Statistical Isotropy (SI) of the universe. For the analysis we use the $\alpha$ statistic that is devised from the contour Minkowski tensor, a tensorial generalization of the scalar Minkowski functional, the contour length. In essence, the $\alpha$ statistic captures the ellipticity of iso-field contours at any chosen threshold value of a smooth random field and provides a measure of anisotropy. 
The SI of the observed convergence map is tested against the suite of realistic simulations of the convergence map  provided by the Planck collaboration.  
We first carry out a global analysis using the full sky data after applying the galactic and point sources mask. We find that the observed data is consistent with SI. 
Further we carry out a local search for departure from SI in small patches of the sky using $\alpha$. This analysis reveals several sky patches which exhibit deviations from simulations with  statistical significance higher than 95\% confidence level (CL). Our analysis indicates  that the source of the anomalous behaviour of most of the outlier patches is inaccurate estimation of noise. We identify two outlier patches which exhibit anomalous behaviour originating from departure from SI at higher than 95\% CL. Most of the anomalous patches are found to be located roughly along the ecliptic plane or in proximity to the ecliptic poles.}

\begin{document}
\maketitle
\flushbottom
\section{Introduction}
\label{sec:intro}

Observations of primary and secondary anisotropies of the cosmic microwave background radiation (CMB henceforth) have unveiled many mysteries about the early universe. CMB photons have been free streaming towards us since the decoupling epoch, where they were last scattered. The temperature and polarization fluctuations of the CMB carry imprints of various critical physical processes that occur in our universe after decoupling, such as the scattering of CMB photons from the free electrons generated during the epoch of reionization, inverse Compton scattering of CMB photons off the high energy electrons in galaxy clusters, and deflection of photon paths by the large scale structure in the universe (i.e. gravitational lensing)~\cite{Blanchard:1987}. 

Out of these sources of secondary anisotropies, gravitational lensing of CMB is of particular interest, as it is a complementary tracer of matter distribution (baryonic and dark matter) in the universe. Furthermore, it has become an important tool in cosmology for studying the growth of structure in the dark matter distribution as well as cosmic acceleration. CMB lensing is an integrated phenomenon that provides a measure of the entire mass distribution, or equivalently, the lensing potential along the line of sight up to the last scattering surface (LSS). As such it is very sensitive to late universe parameters, such as the sum of neutrino masses, the dark energy equation of state and spatial curvature. Lensing remaps the photon distribution in the sky, which leads to the smoothening of the sharp acoustic peaks in the CMB temperature power spectrum ~\cite{Seljak:1995ve, Metcalf:1997ih, Lewis:2006fu}, and generates CMB B-mode polarization at small scales~\cite{Zaldarriaga:1998ar}. CMB lensing was first confirmed in the WMAP data by cross-correlating CMB observations~\cite{Hinshaw:2006ia, Smith:2007rg} with the WISE~\cite{WISE} galaxy data. The Planck 2015~\cite{Ade:2015nch} and 2018~\cite{Aghanim:2018oex} results have provided the most significant $(\sim 40 \sigma)$ detection of gravitational lensing of CMB so far. The data  is further used to reconstruct full-sky lensing potential map and its corresponding angular power spectrum. From the lensing potential one can define the convergence field, which  is related to the lensing potential via the 2-dimensional (2D) Poisson equation. It is interpreted as the sky projection of the surface density of the matter structures that CMB photons encounter in their paths between the LSS and us. The CMB lensing potential provides a new observable, in addition to CMB temperature and polarization anisotropies, to investigate the properties of matter distribution, and hence another way to test our models of cosmological structure formation. The lensing potential estimated from the Planck data maps the mass distribution in the universe over 70\% of the sky up to (z$\sim 1100$) in contrast to the typical surveys  which map the distribution of galaxies, quasars which are concentrated in a limited region of the sky, over limited redshift range. However, observations from different telescopes/surveys are complementary to each other. Ultimately, they all observe the same sky, i.e., the same underlying matter density field. This fact has lead to various cross-correlation studies between CMB lensing and the large scale structure field. For example, there have been joint statistical analyses of the CMB lensing field with the KIDS shear field, Boss galaxies distribution, DESI galaxies~\cite{Hirata:2008ta,Sherwin:2012lq,Pullen:2016glq,Sukhdeep:2016pee,kirk:2016gl} etc. The noise and systematics of CMB lensing observations are quite different from the ones associated with the observations of galaxy density, galaxy lensing or galaxy count, and hence their joint analysis helps in breaking parameter degeneracies and thus improve constrains on some of the cosmological parameters.


The lensing potential or the convergence field, reconstructed from observed lensed CMB, can also be utilized to test the fundamental assumptions of the standard ($\Lambda CDM$) model of cosmology, i.e., the Statistical Isotropy (SI) of the universe on large scales. This assumption of the universe has been directly tested employing different kinds of statistics and cosmological data, for instance, using CMB fields~\cite{Schwarz:2016abs,Planckcol:2016bca,Ghosh:2016tgh}, X-ray background and radio sources~\cite{Alonso:2015ef,Colin:2017gh,Beng:2018bp,Bengaly:2018ba,Rameez:2018ez}, clustering properties of galaxies in large scale structure surveys such as  Wiggle-Z~\cite{WiggleZ:2012} and SDSS~\cite{Ntelis:2017nrj,Sarkar:2018smv}. 


In this paper we test the Statistical Isotropy (SI) of the CMB lensing convergence field using the Tensor Minkowski functionals or Minkowski Tensors (henceforth MTs). MTs are tensor generalizations of scalar Minkowski functionals, defined for excursion sets of smooth random fields. Mathematically they are defined for structures in two- or three- dimensional Euclidean space in ~\cite{Alesker:1999}. A subset of MTs which are translationally invariant in 2D, have been defined for smooth random fields and also generalized to curved spaces  in~\cite{Chingangbam:2017uqv}. MTs carry additional information about the Statistical Isotropy (SI) of the field (interpreted from the orientation/alignment of the structures) and the shape of the structures for each excursion set of a field. One of the translation-invariant rank-2 MTs, the contour Minkowski Tensor (CMT), has been used to test the SI of random fields~\cite{Vidhya:2016,Joby:2018,PKJOBY:2020}. This approach to test the SI of the CMB fields using real space statistics complements harmonic space-based statistics such as BiPolar Spherical Harmonics (BiPoSH)~\cite{Hajian:2004zn,Basak:2006ew,Ghosh:2006xaa}. 

This analysis is a follow-up of our previous paper~\cite{Goyal:2019vkq}, where we carried out theoretical study of the morphology of lensed CMB fields using the CMT on ideal simulations without taking into account the real data complications like masking, noise and residual foregrounds. We quantified the distortions caused by weak lensing in the CMB hotspots and coldspots using the $\beta$ parameter derived from the CMT, which is corresponding to shear. 
Here we test the SI of the reconstructed convergence field from the Planck 2018 data release. In effect, we are probing the SI of the large scale distribution of matter that the CMB photons have interacted with. The assumption of SI of the CMB convergence field should be observationally verified since the detection of violation of SI could have profound implications for cosmology. Recently, Marques et al in~\cite{Marques:2017ejh} investigated SI of the Planck 2015 lensing convergence field using the local variance estimator, and identified sky directions or outlier regions where the weak lensing imprints anomalous signatures to the variance estimator revealed through a $\chi^2$ analyses at a statistically significant level. The goal of this paper is to carry out similar analysis, but using a different methodology. We carry out the global and the local analysis of the convergence field, using $\alpha$ estimator derived from the CMT. We obtained $\alpha$ as a function of the threshold level of the Planck convergence field, and compare with the corresponding $\alpha$ obtained from Planck simulated convergence map. We identify the thresholds where we find more than $2~\sigma$ disagreement between observed data and simulations. We further probe those anomalous thresholds by carrying out our local analysis, by which we mean, the $\alpha$ distribution in  non-overlapping patches in the sky.

This paper is organized as follows. In section~\ref{sec:sec2}, we briefly review the method of the estimation of the convergence field followed by the Planck collaboration. 
In section~\ref{sec:sec3}, we describe the data and the simulations used in our analysis. We also review the definition of the contour Minkowski Tensor and the shape and alignment parameters obtained from it. In section~\ref{sec:sec4}, we present our analysis and the main results. The paper ends with concluding remarks on Section~\ref{sec:sec5}.    

\section{CMB lensing Physics}
\label{sec:sec2}
Weak gravitational lensing (WL) is the  physical process by which CMB photons originating from the LSS are deflected by the gravitational potential of the inhomogeneous distribution of matter along the line of sight. WL of the CMB is an achromatic effect which means that lensing by transverse potential gradients does not change the frequency of the CMB photons. Hence it does
not change the frequency distribution in a given direction, instead it produces an apparent shift in the angular position of the source.

Let $f$ denote a CMB field which can be either temperature, $T$, or the Stoke's parameters $Q,U$ of polarization. Let the deflection caused by lensing in each sky direction, $\hat n$, be denoted by deflection angle  $\vec{d}(\hat n)$. Then the observed CMB field value in some direction $\hat{n}$ can be expressed in terms of the corresponding unlensed field value as, 
\begin{equation}
f^{\rm L}(\hat{n}) = f^{\rm UL}(\hat{n}+\vec{d}).
\end{equation}
 For small perturbations in the linear regime, assuming the Born approximation, the deflection field is given by the gradient of the lensing potential, $\Psi$, as $\vec d = \vec{\nabla}_{\hat{n}}\Psi$. The lensing potential is the projection along the line of sight of the space and time dependent gravitational potential field $\psi$, and  is expressed as,
\begin{equation}
\Psi(\hat{n})=-2\int_0^{\chi^*} {\rm d}\chi \left(\frac{\chi^*-\chi}{\chi\chi^*}\right) \psi \left(\chi\hat{n}; \eta_0-\chi\right),
\end{equation}
where $\chi$ represents the comoving distance, $\chi^*$ is the comoving distance to the LSS, $\psi$ denotes the 3-dimensional gravitational potential at conformal distance $\chi$ along the direction $\hat{n}$, and $\eta_0-\chi$ is the conformal time at which the photon was at position $\chi \hat{n}$. Let the convergence field be denoted by $\kappa$. It is related to the lensing potential via the 2D Poisson's equation as, 
\begin{equation}
\kappa = -\frac{1}{2} {\vec{\nabla}_{\hat{n}}}^2 \Psi.
\end{equation}

Gravitational lensing by large scale structure introduces non-Gaussianity into the CMB and provides a new observable (lensing potential), which can be used as a cosmological probe. Quadratic estimators~\cite{Okamoto:2002,Okamoto:2003kl} have been employed to extract this gravitational lensing signal using CMB modes in harmonic space, incorporating both temperature and polarization data. Configuration space based methods like maximum likelihood method have been proposed as well in~\cite{Hirata:2003}, where authors analyzed the convergence field by using temperature anisotropy in real space.  This lensing potential reconstruction is possible due to higher-order correlations between the multipole moments of CMB fields produced by lensing~\cite{Fber:1997rt, Berf:1997et}. The convergence field is then obtained from the reconstructed lensing potential. The $\kappa$ field captures the magnification or demagnification of CMB hotspots and coldspots caused by gravitational lensing. Another measurable consequence of lensing on CMB is the shearing effect, (area-preserving distortions) on the CMB which may be observable through changes to ellipticity distribution of the hot and cold spot as shown in~\cite{Fber:1997rt}. Unlike the galaxy lensing scenario, where shear is the most crucial lensing observable, for the CMB we can gain information from both shear and magnifications. 

\section{Data and Methodology}
\label{sec:sec3} 
In this section, we describe the set of observed data and simulations that we have used for our analysis, followed by the explaination of the  statistical method used to test for SI of the convergence field.   
\subsection{The Planck convergence map: Observed data and Simulations}
\label{sec:data} 

The Planck 2018 data release ~\cite{Aghanim:2018oex} provides the most significant measurement of the lensing potential and its power spectrum over about $70$ percent of the sky. 

{\em Observed data}: To reconstruct the lensing potential, the method~\cite{Carron:2017} employed by the Planck team are based on quadratic estimators that use the features induced by the lensing process such as the diverse correlations of the CMB temperature (T) and polarization (E and B) modes. The combination of these estimators in a minimum-variance (MV) estimator is used to reconstruct the CMB lensing potential, $\psi^{MV}$. The Planck 2015 lensing estimate was based on CMB temperature and polarization multipoles i.e., $\psi^{TT}, \psi^{EE}, \psi^{TE}, \psi^{TB},\psi^{EB}$ and combination of all these five multipoles in the minimum variance quadratic estimator. However, the Planck 2018 estimate of the lensing potential differs from the previous release because it also includes the contribution from filtered B-modes, $B^{WF}$. The CMB data set used as input to the MV lensing estimator was the foreground cleaned map obtained by passing the raw Planck 2018 full mission frequency maps through the SMICA pipeline. 

The reconstructed lensing potential estimate have much red power spectrum, with most of its power on large angular scales. Cutting the maps with red power spectrum in small portions can cause leakage issues. For this reason, we use in our analysis the convergence map, $\kappa^{WF}$, instead of the lensing potential map. The lensing convergence and its corresponding reconstruction noise, have a much whiter power spectrum specially on large angular scales~\cite{Bucher:2012}.

{\em Simulations}: Next, the simulations we use to perform our analysis are the set of 300 realizations of the convergence field, which constitutes the Planck Full Focal Plane (FFP10) simulations~\cite{Aghanim:2018fcm}. These Monte Carlo realizations comprise of a set of maps that incorporate the dominant instrumental effects 
(detector beam, bandpass and correlated noise properties), scanning (pointing and flags) and data analysis (map-making algorithm and implementation) effects. 

Both the observed and simulation data sets of the convergence field are provided in the form of multipole expansion coefficients, $\kappa_{\ell m}^{MV}$, upto $\ell_{max}=4096$ in HEALPix~\cite{Gorski:2005} \footnote{\url{http://healpix.sourceforge.net/}} FITS format.  
These are related to harmonic coefficients of the lensing potential by,
\begin{equation}
    \kappa_{\ell m}^{MV}= \frac{\ell (\ell +1)}{2} \Psi_{\ell m}^{MV}.
\end{equation}
The $\kappa$ field reconstructed from observed data is highly noise dominated especially at small angular scales. The signal-to-noise ratio is $S/N \approx 1$ at $\ell = 60$.  For this reason, we will construct convergence maps by choosing map resolutions corresponding to the Healpix parameter $N_{\rm side}\le 512$.  The largest multipole value is then chosen to be $\ell_{\rm max} = 2 N_{\rm side}$.   
The maps that are constructed (both observed data and simulations) are then further processed following the steps given below.
\begin{itemize}
\item {\em Bandpass filtering}: We filter out $\ell \leq 8$ modes as done in the Planck 2015 and 2018 lensing analysis. The reason is that the low $\ell$ modes have high sensitivity to the mean-field subtraction. The form of the filter, $F_{\ell}$, is given by the following expression,
\begin{equation}
 \centering
 F_{\ell} = \frac{1}{2} \left\lbrace 1 + {\rm tanh}\left( \frac{\ell - \ell_0}{\Delta} \right) \right\rbrace,
 \label{eqn:filter}
\end{equation}
where $\ell_0$ denotes the center of a chosen band of $\ell$, and $\Delta$ denotes the bandwidth. We choose $\ell_{0} = 8$ and $\Delta = 3$ to carry out our analysis.  
The bandpass filtered multipoles are then generated by multiplying ${\kappa}_{\ell m}$s of the input map with $F_{\ell}$.

\item {\em Wiener filtering}: 
To mitigate the effect of noise, we apply Wiener filter to the spherical harmonic coefficients~\cite{Bobbin:2012ph}, as given by the following expression,
\begin{equation}
    \kappa_{\ell m}^{WF}= \frac{C_{\ell}^{\kappa,fid}}{C_{\ell}^{\kappa,fid}+N_{\ell}^{\kappa}} \kappa_{\ell m}^{MV}.
    \label{eqn:wiener}
\end{equation}
Here $C_{\ell}^{\kappa, fid}$ is the convergence power spectrum in the fiducial cosmological model, and $N_{\ell}^{\kappa}$ is the reconstruction noise power spectrum. 

\item {\em Masking}: We further mask the convergence map using the lens reconstruction analysis mask (lens mask henceforth) provided among the lensing products, which has $f_{\rm sky}=0.67$.  This is an improved mask over the 2015 weak lensing mask due to reduced point source contamination for the same sky fraction.  We downgrade the lens mask to the lower resolution that is compatible with the resolution of the convergence map. 
\end{itemize}

\subsection{Contour Minkowski tensor for smooth random fields and its numerical computation}
 \label{sec:w1_random}
The contour Minkowski tensor (henceforth CMT) is defined for a smooth closed curve, $C$, on a general smooth two-dimensional manifold as,  
\begin{eqnarray}
  {\mathcal W}_1 &\equiv& \int_C \, \hat{T} \otimes \hat{T} \,{\rm d}s, 
\label{eqn:tmf_def}
\end{eqnarray}
where ${\rm d}s$ is the infinitesimal arc length, and $\hat T$ is the unit tangent vector at each point of the curve with the direction chosen to be one of the two possibilities. The symbol $\otimes$ represents the symmetric tensor product of two vectors. 

 Let $f$ denote a random field on the sphere for which we want to compute ${\cal W}_1$. We will work with the mean subtracted field, $u\equiv f-\mu$, where $\mu$ is the mean of $f$.  We have taken field threshold values denoted by $\nu$, and then the set of all points on the manifold where the field has a value greater than or equal to $\nu$ form an excursion set, $\mathcal{Q}_{\nu}$. It consists of connected regions and holes. The boundaries, denoted by  $\partial\mathcal{Q}_{\nu}$, of these connected regions and holes form closed curves. Therefore, each threshold gives a set of curves. 

Let $\nabla u = \left( u_{;1},u_{;2} \right)$ denote the covariant derivative of $u$.  The components of the unit tangent vector, $\hat T$, can be expressed in terms of the field derivatives as, $\hat{T}_i = \epsilon_{ij} \,u_{;j}/\left| \nabla u \right|$, 
 where $\epsilon_{ij}$ is the Levi-Civita tensor in two dimensions. 
 Then, ${\mathcal W}_1$ for the  $k^{th}$ curve in $\partial\mathcal{Q}_{\nu}$ can be expressed as,
 \begin{equation}
   {\mathcal W}_1(\nu,k) = \int_{C_k}  \ds  \,\frac{1}{|\nabla u|^2} \ {\mathcal M}, 
  \label{eqn:w1final}
 \end{equation}
where the matrix $\mathcal{M}$ is given by,
 \begin{equation}
  \mathcal M=  \left(
  \begin{array}{cc} 
    u_{;2}^2 &  -u_{;1} \,u_{;2} \\
      -u_{;1} \,u_{;2} & u_{;1}^2
  \end{array}\right).
  \label{eqn:M}
 \end{equation}
The sum of ${\mathcal W}_1$ (denoted with an overbar) over all curves in $\partial\mathcal{Q}_{\nu}$ is then
 \begin{equation}
   {\overline{\mathcal W}}_1 = \int_{\partial{\mathcal Q}_{\nu}}  \ds  \,\frac{1}{|\nabla u|^2} \ {\mathcal M}.
  \label{eqn:w1_all}
 \end{equation}
 Note that by dividing the above equation by the total area of the manifold under consideration, we can obtain ${\overline{\mathcal W}}_1$ per unit area, which is how the scalar MFs are usually expressed.
 The line integral over the boundary of $\mathcal{Q}_{\nu}$ in eq.~\ref{eqn:w1_all} can be further transformed into an area integral by introducing a Jacobian to give
\begin{eqnarray}
   \overline{\mathcal {W}}_1 &=& \int_{{\mathcal S}^2} \da \, \, \delta(u-\nu)\ \frac{1}{|\nabla u|} \ {\mathcal M},
  \label{eqn:w1_area}
 \end{eqnarray}
where $\da$ is the infinitesimal area element on  ${\mathcal S}^2$ and $\delta(u-\nu)$ is the Dirac delta function. 

 \subsection{Definition and physical meaning of alignment parameter $\alpha$}
 
 Let the eigenvalues of $\overline{\mathcal{W}}_1$ be denoted by $\Lambda_1$ and $\Lambda_2$. Then $\alpha$ is defined to be  the ratio,
\begin{equation}
  \alpha\equiv\frac{\Lambda_1}{\Lambda_2},
\end{equation}
Note that by definition the value of $\alpha$ lies between $0$ and $1$. For a single curve $\alpha$ gives a measure of the ellipticity or anisotropy. Any curve which has $m$-fold symmetry, with $m \ge 3$, has $\alpha=1$~ (see \cite{Chingangbam:2017uqv} for detailed explanation). For a distribution of curves the sum of all  the $\mathcal{W}_1$ matrices defines one overlap curve, whose anisotropy is measured by $\alpha$. Therefore, $\alpha$ measures the relative alignment or the deviation from rotational symmetry in the distribution of curves, and its departure from one gives the degree of anisotropy. 


 When we work with finite resolution maps in the space of compact extent, such as the surface of a sphere,  even for a field which is given to be statistically isotropic, we do not obtain $\alpha$ to be precisely equal to one at each threshold even for a field which is known to be statistically isotropic. The $\alpha=1$ is recovered only in the limit of the total perimeter tending to infinity. At threshold values close to zero, where the perimeter of $\mathcal{Q}_{\nu}$ is the largest, $\alpha$ is usually closest to one. For a random distribution of a few curves, the probability that they will be isotropically distributed is small.  As a result, at higher $|\nu|$, the values of $\alpha$ decrease from unity. Therefore, it is essential to take into account the threshold dependence of $\alpha$ when probing statistical isotropy of finite resolution fields. 
 
\subsubsection{Estimator for $\overline{\mathcal{W}}_1$ and $\alpha$}

In order to numerically calculate Eq.~\eqref{eqn:w1_area} we require, first, threshold bins, and secondly we need to express the integral as a sum over pixels on the sphere. Let  $\Delta \nu$ denote threshold bin size. Then the $\delta-$function can be approximated as (~\cite{Schmalzing:1997uc,Chingangbam:2017uqv}),
 \begin{equation}
   \delta \left( u-\nu \right) = 
\left\{ \begin{array}{l}
    \frac{1}{\Delta \nu}, \quad {\rm if} \ 
   u \in \left( \nu-\frac{\Delta \nu}{2} , \nu+\frac{\Delta \nu}{2} \right)\\
    0, \quad {\rm otherwise}. 
  \end{array}
  \right.
  \label{eqn:delta}
 \end{equation}
 For the pixellized sky the total number of pixels is given by $N_{\rm pix}=12N_{\rm side}^2$.  The estimator for  $\overline{\mathcal {W}}_1$ (given by eq.~\ref{eqn:w1_area}) for masked pixellized sky can then be expressed as
\begin{eqnarray}
   \overline{\mathcal {W}}_1 &=& \frac{1}{N}\sum_{i=0}^{N_{\rm pix}} \, \frac{w_i}{\Delta\nu}  \ {\mathcal I}_i,
  \label{eqn:w1_discrete}
 \end{eqnarray}
 where ${\mathcal I}_i \equiv {\mathcal M}/|\nabla u|$, with the right hand side evaluated at pixel $i$, and ${\mathcal M}$ is given by eq.~\ref{eqn:M}. The variable $w_i$ is the pixel weight which has value zero at pixels that are masked, and one otherwise. $N$ is the count of all pixels having $w_i=1$. 
 
 By diagonalizing $\overline{\mathcal {W}}_1$ we get its eigenvalues and obtain $\alpha$ by taking their ratio.  This method has inherent numerical error coming from the discrete approximation of the delta function~\cite{Lim:2011kd}. However, it was shown in~\cite{Goyal:2019vkq} that the error mostly cancels out when computing $\alpha$. and hence 
is well suited for applications where $\alpha$ is used to extract physical information.

\section{Analysis and results}
\label{sec:sec4} 

In this section we present our analysis of SI of the Planck convergence map using the $\alpha$ statistic. In order to keep the terminology clear we will refer to the convergence map obtained from Planck data as the {\em observed convergence map}. We first rescale the convergence field by its variance $\sigma_0$. This makes the rescaled field threshold range to be of order one.  
Then  we compute $\overline{\mathcal {W}}_1$, and from it $\alpha$, for 33 equally spaced threshold bins, of width $\Delta\nu=0.25$,  ranging from $-4.0 \leq \nu \leq 4.0$. 

{\em Minimizing numerical error due to mask boundaries}: In order  to minimize numerical error near sharp mask boundaries that can arise due to harmonic transforms, we first mask the field using downgraded lens mask which is apodized using the same smoothing kernel as the convergence maps. 
Then, we include only those pixels which are sufficiently far away from the boundary in the calculation of $\alpha$. This is done using a parameter $s_m$, whose value lie between zero and one, and only pixels of the field for which the corresponding smoothed lens mask has values greater than $s_m$ are included. Since, as $s_m$ increases towards one, the fraction of included pixels decreases, and hence the statistical significance of the results will decrease. Therefore, it is best to select an optimum value of $s_m$ such that the numerical error is minimized and the statistical significance is maximized. A rough estimate shows that a smoothed mask pixel value greater than $0.89$ roughly corresponds to $>2\theta_s$ distance from the mask boundary, $\theta_s$ being the smoothing scale. Hence we use $s_m=0.89$ for our calculations.

\subsection{Understanding the effects of bandpass filter and masking on $\alpha$ using ideal simulations}
\label{sec:ideal}

Before analysing the observed data set and looking for subtle signatures of departure from SI in the universe, it is important to have a clear understanding of the effects of masking and bandpass filtering on $\alpha$ using ideal Gaussian isotropic maps of the convergence field. This step is important to isolate any physical phenomenon from the systematic effects. For this purpose we simulate 300 Gaussian isotropic realizations of $\kappa$ and then compute $\alpha$ using the method outlined in section \ref{sec:w1_random}. The input lensing potential or convergence power spectrum is obtained from \texttt{CAMB}~\cite{Lewis:1999b}\footnote{\url{https://camb.info/}}. The values of the input $\Lambda$CDM parameters are: $\Omega_bh^2=0.02216$, $\Omega_{cdm}h^2= 0.1203$, $H_0=67\, {\rm km\,s^{-1}\, Mpc^{-1}}$, $\tau=0.06$, $n_s=0.964$, $A_s=2.119 \times 10^{-9}$, where the symbols have their usual meanings. These are the same fiducial cosmological parameters which are used as input for the Planck convergence simulations.

\begin{figure}
\centering
\includegraphics[scale=0.4]{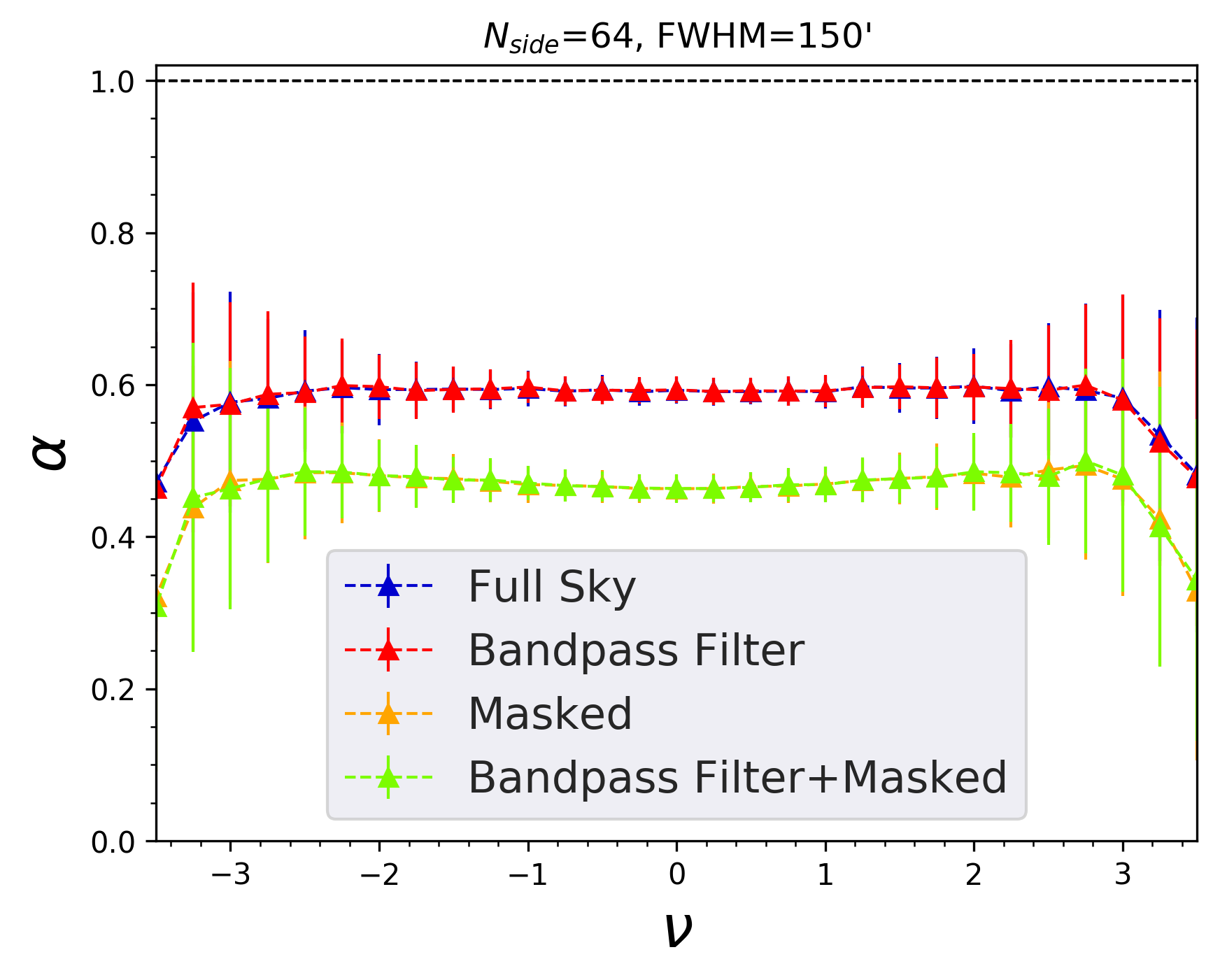} \quad
\includegraphics[scale=0.4]{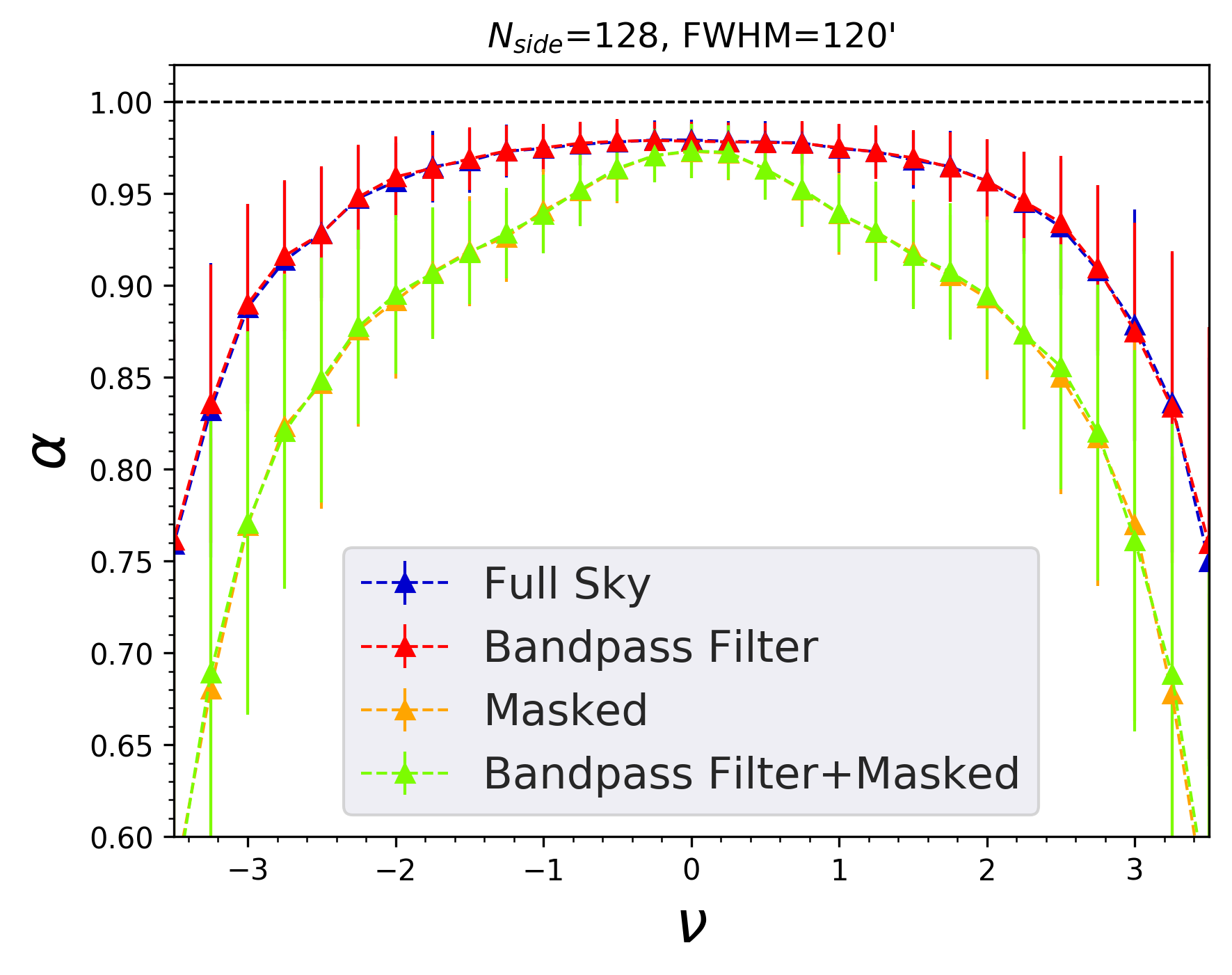}\\
\includegraphics[scale=0.4]{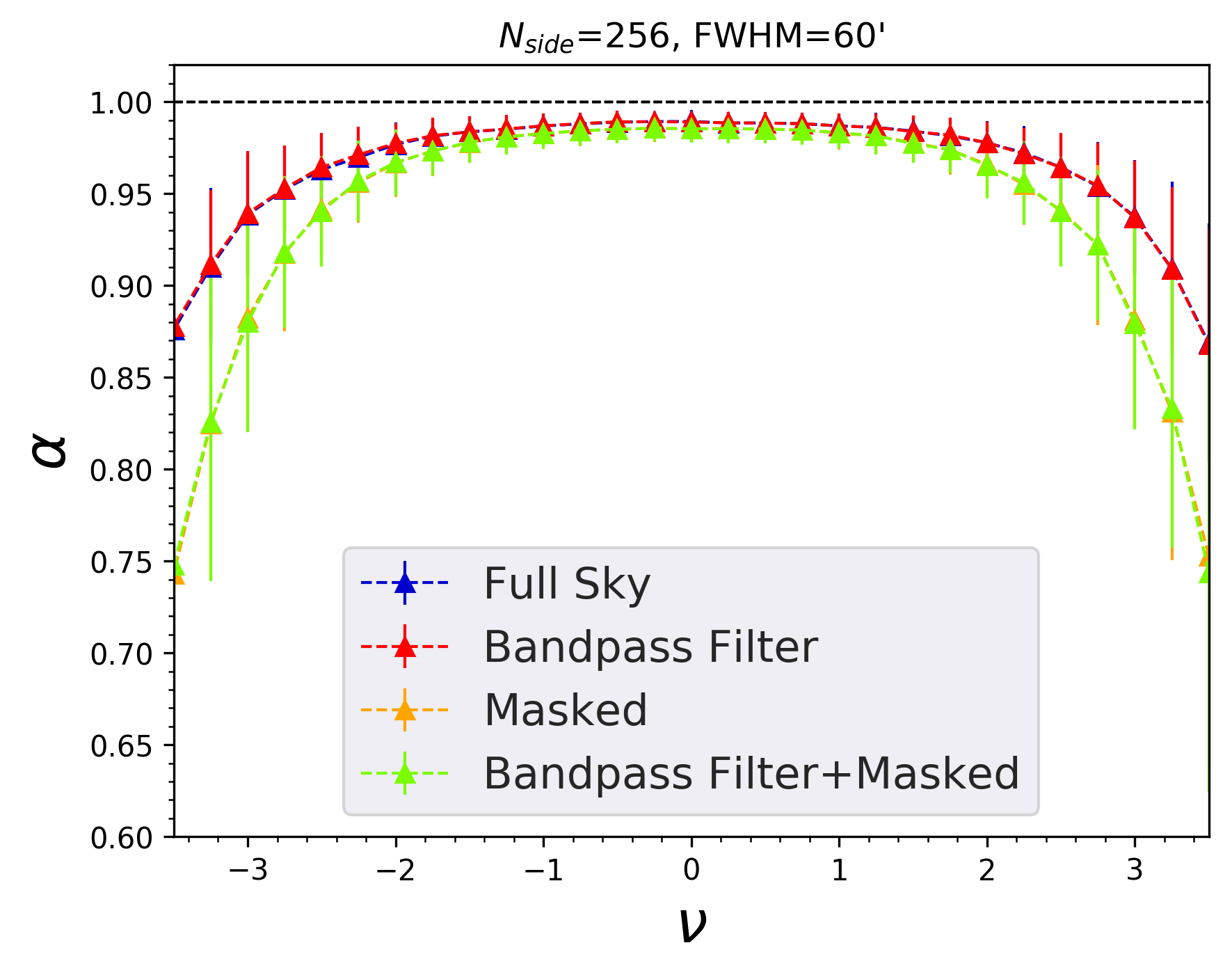} \quad 
\includegraphics[scale=0.4]{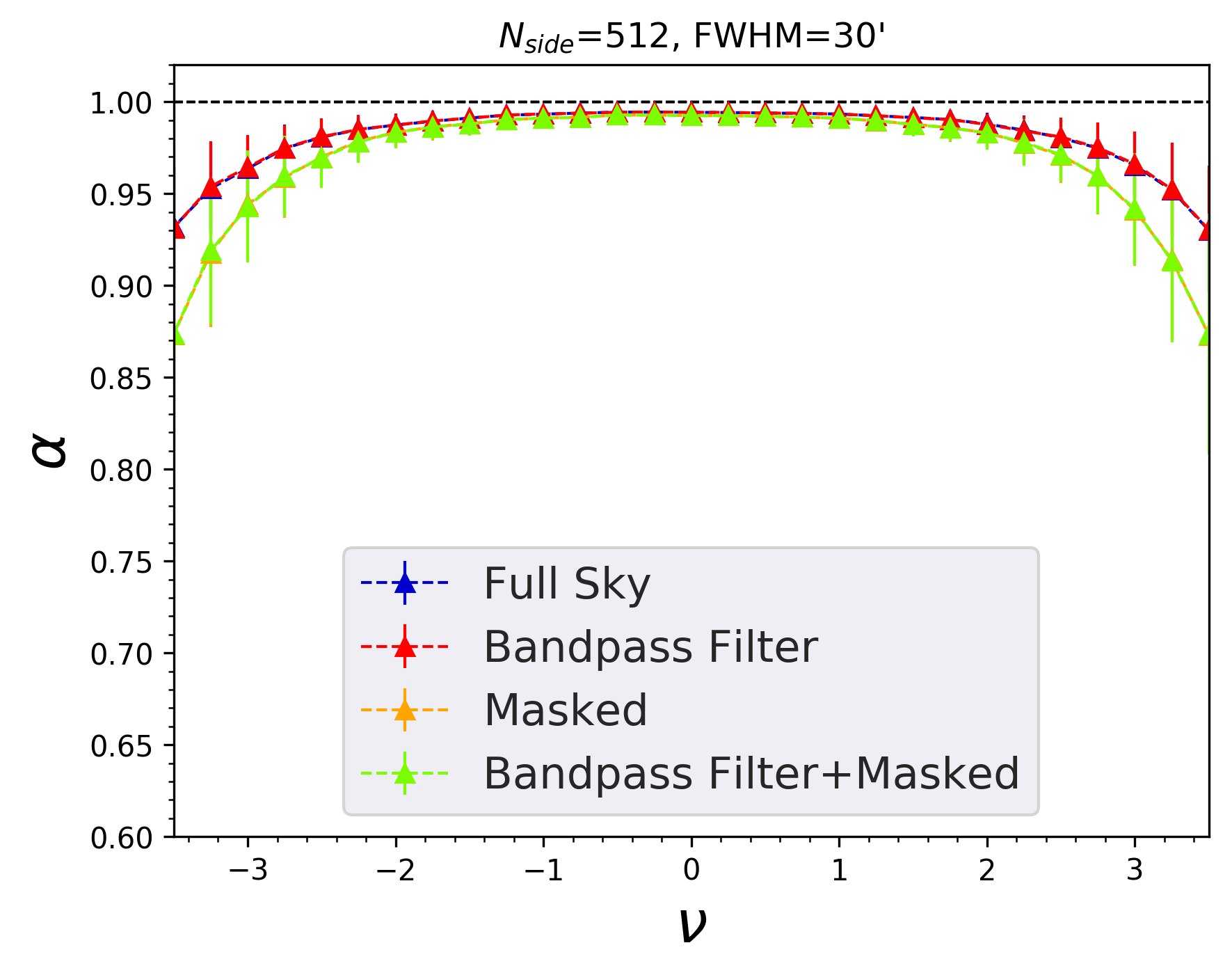}
\caption{Mean values of $\alpha$ and standard deviation obtained from isotropic Gaussian simulations of the convergence map for different resolutions $N_{\rm side}=64,\ 128,\ 256$ and 512. Different colors here represent $\alpha$ calculated for  full sky (red), masked (green), bandpass filtered (yellow), and masked plus band pass filtered (blue).}
\label{fig:fig1}
\end{figure}

We compare $\alpha$ as a function of the threshold for the full sky case with the corresponding $\alpha$ obtained after (a) applying lens mask with $s_m =0.98$, (b) no masking but band bass filtering, and (c) band pass filtering and masking. In figure~\ref{fig:fig1} we have shown $\alpha$ for these three different cases. 
The four panels correspond to four different resolutions given by $N_{\rm side}=64,\ 128, \, 256$ and  $512$. The red curves correspond to the full sky case, while the green curves represent the masked case, yellow curves are for the bandpass filtered case with cut-off scale $\ell_0=8$, and the blue curves represents the alpha for band pass filtered and masked maps. For the ideal case, we expect $\alpha$ to be close to one at all thresholds considered here. This is what we find that the ideal convergence field is isotropic as $\alpha$ is close to one, which is as expected from the standard model of cosmology. Applying lens mask to the ideal $\kappa$ field, reduces the number of structures in the field which leads to a drop in the value of $\alpha$ at large thresholds. This effect is more prominent at lower resolutions i.e. $N_{side}=64, 128$ where already the ideal field has few structures. We also note that the bandpass filtering does not have much effect on the value of $\alpha$.

\subsection{Quantifying the statistical significance of $\alpha$ and error bars}
\label{sec:chi}

The $\alpha$ statistic follows the Beta distribution given by the following expression,
$$P(\alpha)= \frac{\Gamma(a+b)}{\Gamma(a)\Gamma(b)}\alpha^{a-1} (1-\alpha)^{b-1},$$ where $a > 0,\ b> 0$ are parameters that depend on the cosmological model~\cite{Prava:2021}. Hence the standard deviation of $\alpha$, which we denote by $\sigma_{\alpha}$, will not be equivalent to the usual 68\% confidence interval.  
So, we first reconstruct the probability density function (PDF) of $\alpha$ using the 300 FFP10 realizations and using it determine the confidence intervals as described below.   
Let $\widetilde\alpha^{\,\rm sim}(\nu)$ denote the median value of $\alpha$ at each threshold obtained from the simulations. Let the ranges of $\alpha$ be given by $\alpha_a=\left(\widetilde\alpha^{\,\rm sim}-\delta^{(j)}_{-}\right)$ to $\alpha_b= \left(\widetilde\alpha^{\,\rm sim}+\delta^{(j)}_{+}\right)$, which denotes the confidence intervals such that $j=1,2,3,\ldots$ correspond to 68\%,  95\% and 99\%, and so on, respectively.   Then we determine $\delta^{(j)}_{-}$ and $ \delta^{(j)}_{+}$ from the condition 
\begin{equation}
\int_{\widetilde\alpha^{\,\rm sim} -\delta^{(j)}_-}^{\widetilde\alpha^{\,\rm sim}} {\rm d}\alpha \,P(\alpha) = \int_{\widetilde\alpha^{\,\rm sim}}^{\widetilde\alpha^{\,\rm sim}+\delta^{(j)}_+} {\rm d}\alpha \,P(\alpha) = p/2,
\end{equation}
where $p=0.68,0.95,0.99$ for $j=1,2,3$. We find that the difference between the mean and median values of $\alpha$ is actually quite small in all cases. They differ at most by 1\% at $|\nu| \sim 3$ and this difference gets smaller towards $|\nu| \sim 0$.  
\begin{figure}
\centering
\includegraphics[scale=0.3]{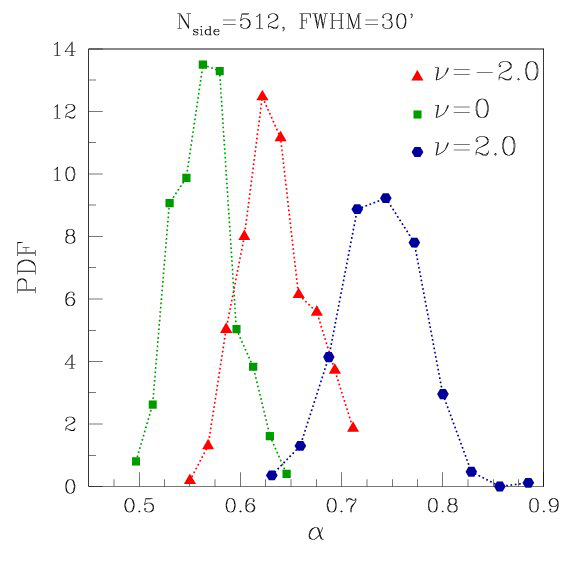}
\includegraphics[scale=0.3]{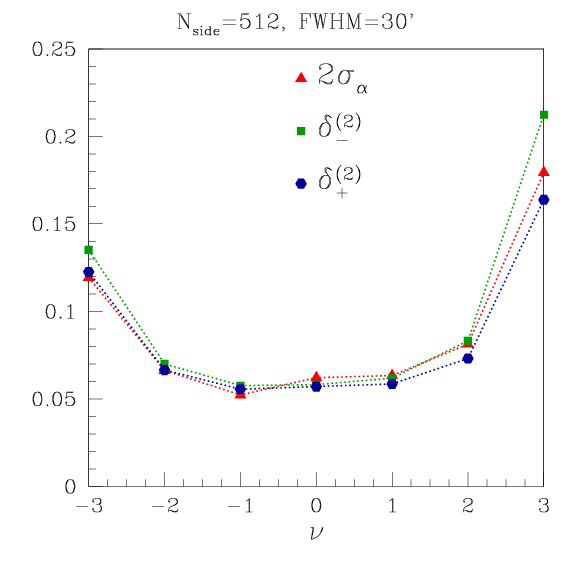}
\caption{{\em Left}: $P(\alpha)$ for some selected values of $\nu$ for $N_{\rm side}=512$. {\em Right}: $2\sigma_{\alpha}$ (red triangle), $\delta^{(2)}_-$ (green square) and  $\delta^{(2)}_+$ (blue circle), for the same $N_{\rm side}$ as the left panel.}
	\label{fig:figp}
\end{figure} 

To demonstrate, the left panel of figure~\ref{fig:figp} shows the reconstructed  $P(\alpha)$ for some selected $\nu$ values, for $N_{\rm side}=512$. 
We observe that the maxima of the PDF for $\nu=0$ occurs at smaller value of $\alpha$ compared to $\nu=2$. Moreover, there is no symmetry between positive and negative $\nu$. These are due to the presence  of instrumental noise, foreground residuals and other systematic effects. The panel on the right shows  comparison of $2\sigma_{\alpha}$ with  $\delta^{(2)}_-$ and  $\delta^{(2)}_+$, calculated using $p=0.95$. 
The values of the three quantities do not differ much. There is mild mismatch at higher $|\nu|$ values where  $2\sigma_{\alpha}$ slightly underestimates $\delta^{(2)}_-$ and overestimates $\delta^{(2)}_+$.  
We find similar behaviour for different values of  $N_{\rm side}$. 

Let $\alpha^{\rm obs}$ denote the value of  $\alpha$  for the observed data, and let $\widetilde\alpha^{\rm sim}$ mean/median $\alpha$ value from 300 FFP10 simulations, then their difference $\Delta\alpha$ is given by, 
\begin{equation}
\Delta\alpha \equiv \alpha^{\rm obs} - \widetilde\alpha^{\rm sim}.
\end{equation}
Then in order to quantify the statistical significance of $\Delta\alpha$ at each threshold we define 
the variable $\widetilde\chi^{(j)}$ as follows\footnote{$|\widetilde\chi^{(1)}|$ reduces to the square root of the standard chi-squared statistic for Gaussian distribution.}:
\begin{equation}
 \widetilde \chi^{(j)} = \left\{
  \begin{array}{c}
 \Delta\alpha/\delta^{(j)}_-, \quad {\rm if \ \Delta\alpha<0}, \\
        \Delta\alpha/\delta^{(j)}_+, \quad {\rm if \ \Delta\alpha>0}.
\end{array} \right.
\end{equation}  
 If $|\widetilde\chi^{(j)}|>1$ then it implies that $\alpha^{\rm obs}$ is outside the confidence interval corresponding to $j$. 
 Since there can be unknown systematic or  physical effects at different threshold levels we choose to interpret $\widetilde\chi^{(j)}$ as a function of threshold, rather than condense the information of the threshold variation into a single value. It can be shown that the values of $\alpha$ at neighbouring thresholds are uncorrelated if the threshold bin size is sufficiently large. Our choice of $\Delta\nu=0.25$ is sufficiently large and hence $\delta^{(j)}_-$ and $\delta^{(j)}_+$ in the denominator of $\widetilde\chi^{(j)}$ captures the full covariance information.
 
 An important point to keep in mind when we interpret our results is the following. Since lower value of $\alpha$ means higher degree of alignment of structures, a significantly negative value of $\Delta\alpha$, and consequently of  $\widetilde\chi^{(j)}$,  implies that the observed data has higher level of   anisotropy in comparison to the expectation from the median value of $\widetilde\alpha^{\rm sim}$. On the other hand,  as seen in  section~\ref{sec:ideal}, $\alpha$ value for a generic field depends on the number of structures and is larger for field with higher number of structures. Therefore, a significantly large positive value of $\Delta\alpha$ indicates anomalous behaviour, but does not indicate  higher level of anisotropy. 

Equipped with the understanding of the effects of masking and bandpass filtering on the ideal convergence field, and the statistics of $\alpha$, we now focus on our stated goal of testing SI of the observed convergence map and the corresponding FFP10 simulations. 
In the following we perform our analysis using two complementary approaches: (1) {\em global analysis}, where  we calculate $\alpha$ for the masked full sky convergence maps, and (2) {\em local analysis}, where we calculate $\alpha$ for non-overlapping small patches of the sky.

\subsection{Global analysis}
\label{sec:global}

For the global analysis we first construct the maps of the observed and FFP10 simulated convergence field following the steps outlined in section~\ref{sec:data}. 
For the masking we use the conservative value of $s_m=0.98$. 
The sky fraction for the full resolution $(N_{\rm side}=2048$) mask is $f_{sky}=0.67$. After apodizing the lens mask and then applying the $s_m$ condition the sky fractions become  $f_{sky}=$ 0.48, 0.49, 0.53 and 0.60 for $N_{\rm side}$ = 64, 128, 256 and 512, respectively. Then $\alpha$ is calculated from each of the maps constructed as above. We will discuss the results for the cases  with and without Wiener filtering. 

\begin{figure}[ht] 
\centering
\includegraphics[scale=0.35]{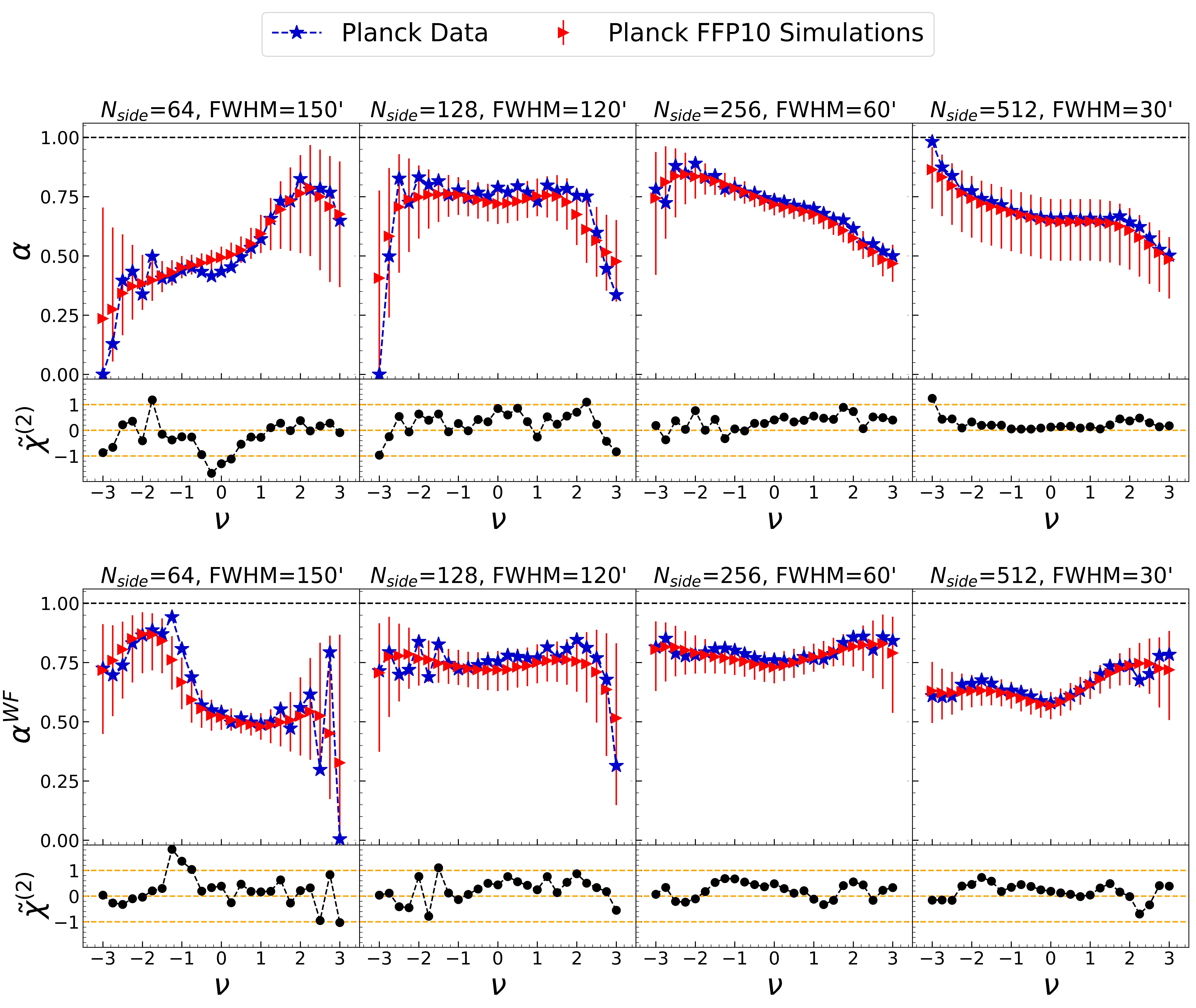}  
\caption{{\em Top row}: Upper panels show $\alpha$ while lower panels show  $\widetilde\chi^{(2)}$ for the Planck convergence map (blue stars) and FFP10 simulations (red triangles), without Wiener filtering, for different values of $N_{\rm side}$. {\em Bottom row}: Same as first row but for Wiener filtered maps.}
	\label{fig:fig3}
\end{figure}

We first discuss the results for the case without Wiener filtering.  The $\alpha$ and corresponding $\widetilde\chi^{(2)}$ values as functions of the field threshold are presented in the top row of figure.~\ref{fig:fig3}. The four panels correspond to $N_{\rm side}=$\,64,\, 128,\,256 and 512, which are the same as in figure~\ref{fig:fig1} with the same smoothing FWHM values. The $\alpha$ values computed from the observed $\kappa$ map are represented by blue stars, while the median value and error bars given by $\delta^{(2)}_-$ (lower) and $\delta^{(2)}_+$ (upper)  are shown in  red.  The black curves in the lower panels show $\widetilde\chi^{(2)}$. We observe the following points:
\begin{itemize}
\item We find that $\alpha$ values differ strongly in terms of amplitude as well as shape of the threshold dependence, between the FFP10 simulations and the ideal cases shown in figure~\ref{fig:fig1}. We observe a dip in the $\alpha$ curve at around $\nu=0$. Also, we note that $\alpha$ is asymmetric about $\nu=0$, unlike the case of ideal simulations. We find that the curve slopes down towards negative threshold values for $N_{\rm side}=64$, while for higher resolution cases the downward slope is towards positive thresholds. This difference can be attributed to the complex noise (arising from both instrument and lensing reconstruction) properties contributing in the observed data and the corresponding simulations.

\item From the values of  $\widetilde\chi^{(2)}$ we infer that there is good agreement between the observed data and FFP10 simulations for all the four cases except for some threshold values, particularly for $N_{\rm side}=64$.  
\end{itemize}

Next we discuss the results for the case with Wiener filtering. In the bottom row panels of figure~\ref{fig:fig3} we show $\alpha$ for the same observed data (blue stars) and simulations (red triangles) as in the top row. The superscript `WF' refers to Wiener filter. Wiener filtering optimally weighs for the noise and hence, is expected to suppress the modes with large noise contribution. We observe some changes in the results after applying Wiener filter. For $N_{\rm side}=64$ we observe that the $\alpha$ curve now is tilted toward the positive threshold values in contrast to the without Wiener filtering case. While for $N_{\rm side}=256, 512$ we find that $\alpha$ becomes  relatively  more symmetric about $\nu=0$, as the (large $\ell$ values) modes with high noise contribution have been suppressed by the Wiener filter. Even though, the shape of the $\alpha$ curve obtained here also do not mimic the $\alpha$ curve in the case of ideal Gaussian isotropic $\kappa$ simulation, the observed $\kappa$ is consistent with Planck isotropic  simulations within $2 \sigma$. Agreement between the corresponding values computed from observations and simulations imply that the observed maps are statistically isotropic. We also observe that  $\widetilde\chi^{(2)}$ has positive values at most of the thresholds for almost all the $N_{\rm side}$ values considered here, in both with and without Wiener filtering case.  positive values of $\widetilde\chi^{(2)}$ indicate  presence of higher number of structures in the observed $\kappa$ field than the corresponding simulated field.

We are guided by two factors to complement the global analysis  by further analysis after restricting to smaller sky patches. 
The first is that any anomaly in localized sky regions can get washed out when $\alpha$ is computed over larger regions. Secondly, we do see some hints of anomaly in the global analysis, and these can get sharpened when analyzed over smaller regions. Therefore, we expect the statistical significance of the  positive values of $\widetilde\chi^{(2)}$ to get enhanced. 


\subsection{Local patch analysis}

We now focus on analysis of small patches of the sky. We first pixellize the sky using a low value of the Healpix resolution parameter which we denote  by  $N_{\rm side}^{\rm local}$. Each pixel determined by  $N_{\rm side}^{\rm local}$ is  referred to as a {\em patch}. We will work with $N_{\rm side}^{\rm local}$ = 2 and 4. For $N_{\rm side}^{\rm local}$ = 2 we get 48 patches, each patch having angular size $(29.3^{\circ})^2$. For $N_{\rm side}^{\rm local}$ = 4 there are 192 patches and each patch has angular size $(14.7^{\circ})^2$. We identify each patch by its pixel number in the `ring' pixel numbering format in Healpix. 
Next, we pixellize the sky using a higher $N_{\rm side}$ value which we denote with a superscript `global', by  $N_{\rm side}^{\rm global}$. The number of pixels in each patch defined above is then given by $N_{\rm pix}^p=(N_{\rm side}^{\rm global}/N_{\rm side}^{\rm local})^2$.
We work with $N_{\rm side}^{\rm global}=128$ and 256. In each case we use the maximum multipole value of $\kappa$ given by $\ell_{\rm max} \sim 2N_{\rm side}^{\rm global}$. Therefore, the physical information encoded in the maps constructed with $N_{\rm side}^{\rm global}=128$ is contained in maps having $N_{\rm side}^{\rm global}=256$, but the converse is not true. The values of $N_{\rm side}^{\rm local}$ and $N_{\rm side}^{\rm global}$ have been chosen keeping in mind that the SNR of the observed convergence map becomes increasingly smaller at high  multipoles, and that the patches should have good enough resolution for statistical analysis using morphological properties.  
Table~\ref{tab:table1} summarizes the values of $N_{\rm side}^{\rm local}$  and $N_{\rm side}^{\rm global}$ considered in our analysis, along with the relevant numbers of patches and pixels, and the angular sizes of patches.

\renewcommand{\arraystretch}{1.2}
\begin{table}
\begin{center}
\begin{tabular}{||c|p{1.9cm}|p{2.5cm}|c|p{2.5cm}||}   
\hline\hline
$N_{\rm side}^{\rm local}$ & Total no of patches & Angular size of a patch & $N_{\rm side}^{\rm global}$  & No of pixels in each patch \\ 
\hline\hline
\multirow{2}{*}{2} &   \multirow{2}{*}{\hskip .7cm 48}  & \multirow{2}{*}{\hskip 0.5cm $(29.3^{\circ})^2$ }  &
\multirow{1}{*}{128} &  \hskip .8cm  4096  \\
 \cline{4-5}  
&&& \multirow{1}{*}{256}  &  \hskip .6cm 16384 \\
\hline\hline
\multirow{2}{*}{4} &    \multirow{2}{*}{\hskip .5cm 192}  & \multirow{2}{*}{\hskip 0.5cm $(14.7^{\circ})^2$ }  &
\multirow{1}{*}{128} &  \hskip .8cm  1024  
\\ \cline{4-5} 
&&& \multirow{1}{*}{256}  &  \hskip .8cm 4096   \\ 
\hline\hline
  \end{tabular}
  \caption{Table showing the numbers of patches and their angular size. and number of  pixels in each patch, for the values of $N_{\rm side}^{\rm global}$ and $N_{\rm side}^{\rm local}$ considered in our local analysis. }
  \label{tab:table1}
  \end{center}
\end{table}

\begin{figure}[H]
	\begin{centering}	
			\includegraphics[height=1.7in,width=1.8in]{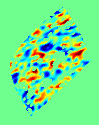} 
			\hskip 1cm	
		{\includegraphics[height=1.9in,width=3.7in]{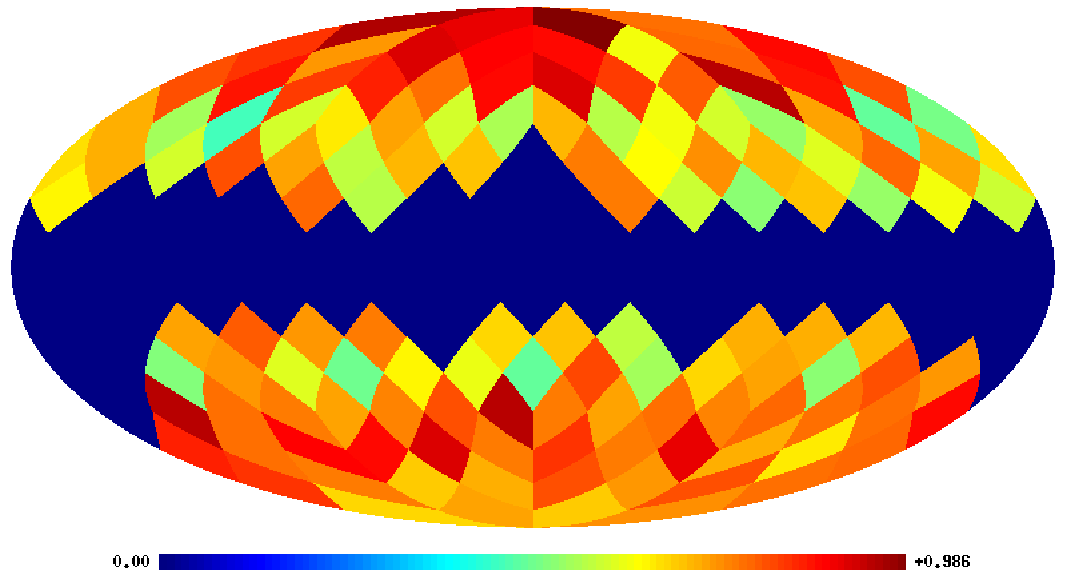}}
\caption{{\em Left}: An example of a patch of the $\kappa$ map. {\em Right}: Map showing values of $\alpha$ for sky patches at $\nu=-0.5$, with each patch being one of the 192 pixels given by $N_{\rm side}^{\rm local}=4$.   The value of $N_{\rm side}^{\rm global}$  is 256 and the valid pixel fraction is $p_{\rm frac}= 0.6$.} 
		\label{fig:fig4}
	\end{centering}
\end{figure}

The estimator of $\overline{\mathcal W}_1$, which for the global analysis is given by eqn.~\ref{eqn:w1_discrete}, is modified for each patch indexed by $p$, to the following form,
\begin{eqnarray}
   \overline{\mathcal {W}}_{1,p} &=& \frac{1}{N_p}\sum_{i=0}^{N_{\rm pix}} \, \frac{w_{p,i}}{\Delta\nu}  \ {\mathcal I}_i,
  \label{eqn:w1_patch}
 \end{eqnarray}
 where the weight $w_{p,i}$ has one if the pixel belongs to the $p^{\rm th}$ patch {\em and} if the apodized lens mask at that pixel has value $>s_m$, else it has value zero. The normalization $N_p$ is now the count of pixels in each patch having $w_{p,i}=1$.   
We also exclude patches that have high percentage of masked regions. This is done by calculating the ratio, $p_{\rm frac}$, of  the number of  pixels that are not masked ({\em valid} pixels) to the total number of pixels in each patch. We choose $p_{\rm frac}=0.6$. 
Figure~\ref{fig:fig4} visually demonstrates the local analysis. The left panels shows  the $\kappa$ field in one patch. The right panel shows a map  of $\alpha$ values for each patch  computed from the observed convergence map,  at threshold $\nu=-0.5$ for the case of $N_{\rm side}^{\rm local}=4$ and $N_{\rm side}^{\rm global}=128$. 

{\em Identification of outlier patches}: 
We calculate $\Delta\alpha$ for each patch  at each threshold. Then we compute  $\widetilde\chi^{(j)}(\nu)$, $j=1,2,3,...$. 
At high thresholds the excursion sets for a typical patch consist of few structures and hence the analysis of statistical significance can become unreliable. Therefore, we restrict the threshold range for identifying outlier patches to the conservative range  $-2\le \nu\le 2$.  
If $|\widetilde\chi^{(2)}(\nu)| >1$ (equivalent to higher than $2\sigma$)
we consider the patch to be anomalous at that threshold.  
A patch can exhibit anomaly at one or more threshold values.
To identify a patch as an {\em outlier}  for each value of  $N_{\rm side}^{\rm local}$ we demand  that it satisfies {\em either} of the following two conditions:
\begin{enumerate}
\item It must show anomaly for both $N_{\rm side}^{\rm global}$ = 128 and 256. This condition ensures that anomalous behaviour for $N_{\rm side}^{\rm global}$ = 128 is also manifested in the higher resolution case,  and that the anomaly is robust against variation of resolution. 

\item For patches that are not common as above they must be anomalous at three or more threshold values. This condition ensures that the anomaly is not just a  statistical fluctuation.  
\end{enumerate}
The outlier patches that have been identified are listed in Table~\ref{tab:table2}. 
Patches shown in magenta have positive value of $\widetilde\chi^{(2)}(\nu)$, while those in blue have negative value of $\widetilde\chi^{(2)}(\nu)$ at all anomalous thresholds. 
  The patches that are {\em not} common between $N_{\rm side}^{\rm global}$=128 and 256 are highlighted by the black boxes.

\renewcommand{\arraystretch}{1.7}
\begin{table}
\begin{center}
\begin{tabular}{|c|c|p{1.9cm}|p{1.2cm}|p{7.5cm}|}   
\hline\hline
$N_{\rm side}^{\rm local}$&  $N_{\rm side}^{\rm global}$ &  No of valid patches 
& No of outlier patches 
& \hskip 1.2cm Patch ids of outlier patches\\ 
\hline\hline
\multirow{2}{*}{2} &   
\multirow{1}{*}{128} &     \hskip .6cm  28 & \hskip .5cm 8 &    \textcolor{magenta}{\bf 11, 17, 36, 37,  41, 42, 43, 45}  \\
\cline{2-5}  
 &   \multirow{1.7}{*}{256} &     \multirow{1.7}{*}{\hskip .6cm  29} &  \multirow{1.7}{*}{\hskip .4cm 11} &    \textcolor{magenta}{\bf  11,  17,  36,  37}, \fbox{\textcolor{magenta}{\bf 38}},  \fbox{\textcolor{magenta}{\bf 40}}, \textcolor{magenta}{\bf  41,  42,  43,  45}, \fbox{\textcolor{magenta}{\bf 46}}   
 \\ 
\hline\hline
\multirow{4.5}{*}{4}   &   
\multirow{2.4}{*}{128} &   \multirow{2.4}{*}{\hskip .5cm  123}  & \multirow{2.4}{*}{\hskip .3cm  22} & \textcolor{blue}{\bf  0}, \textcolor{magenta}{\bf 3}, 
\textcolor{blue}{\bf 29}, \textcolor{magenta}{\bf  30,  31},
\textcolor{magenta}{\bf 44,  45,  53,  55,  65}, \fbox{\textcolor{magenta}{\bf 67}}, \fbox{\textcolor{magenta}{\bf 134}}, \fbox{\textcolor{magenta}{\bf 140}}, \textcolor{blue}{\bf 142}, 
\textcolor{magenta}{\bf  156}, \textcolor{blue}{\bf  161}, \textcolor{magenta}{\bf 162, 163, 165, 
172}, \fbox{\textcolor{blue}{\bf 179}}, \textcolor{magenta}{\bf  184}  
\\ \cline{2-5} 
&   \multirow{2.3}{*}{256}  &   \multirow{2.3}{*}{\hskip .5cm  121} & \multirow{2.3}{*}{\hskip .3cm  23}  & \textcolor{magenta}{\bf 0, 3}, \textcolor{blue}{\bf 29, 30}, \textcolor{magenta}{\bf 31},  \textcolor{magenta}{\bf 44, 45},  \textcolor{magenta}{\bf 53, 55},  \textcolor{magenta}{\bf 65},  \fbox{\textcolor{magenta}{\bf 125}}, \textcolor{magenta}{\bf 140, 142}, \fbox{\textcolor{magenta}{\bf 152}}  \textcolor{magenta}{\bf 156}, \fbox{\textcolor{magenta}{\bf 158}}, \fbox{\textcolor{magenta}{\bf 160}}, \textcolor{magenta}{\bf 161, 162, 163},  \textcolor{magenta}{\bf 165},  \textcolor{magenta}{\bf 172},  \textcolor{magenta}{\bf 184}  
 \\ \hline\hline
  \end{tabular}
  \caption{Table showing the numbers of anomalous patches, and their identification numbers. Patches shown in pink have positive value of $\widetilde\chi^{(2)}(\nu)$, while those in blue have negative values of $\widetilde\chi^{(2)}(\nu)$ at all anomalous thresholds. 
  The patches that are {\em not} common between $N_{\rm side}^{\rm global}$=128 and 256 are highlighted by the black boxes.} 
  \label{tab:table2}
  \end{center}
\end{table}

\begin{figure}[ht] 
\centering
\includegraphics[scale=0.3]{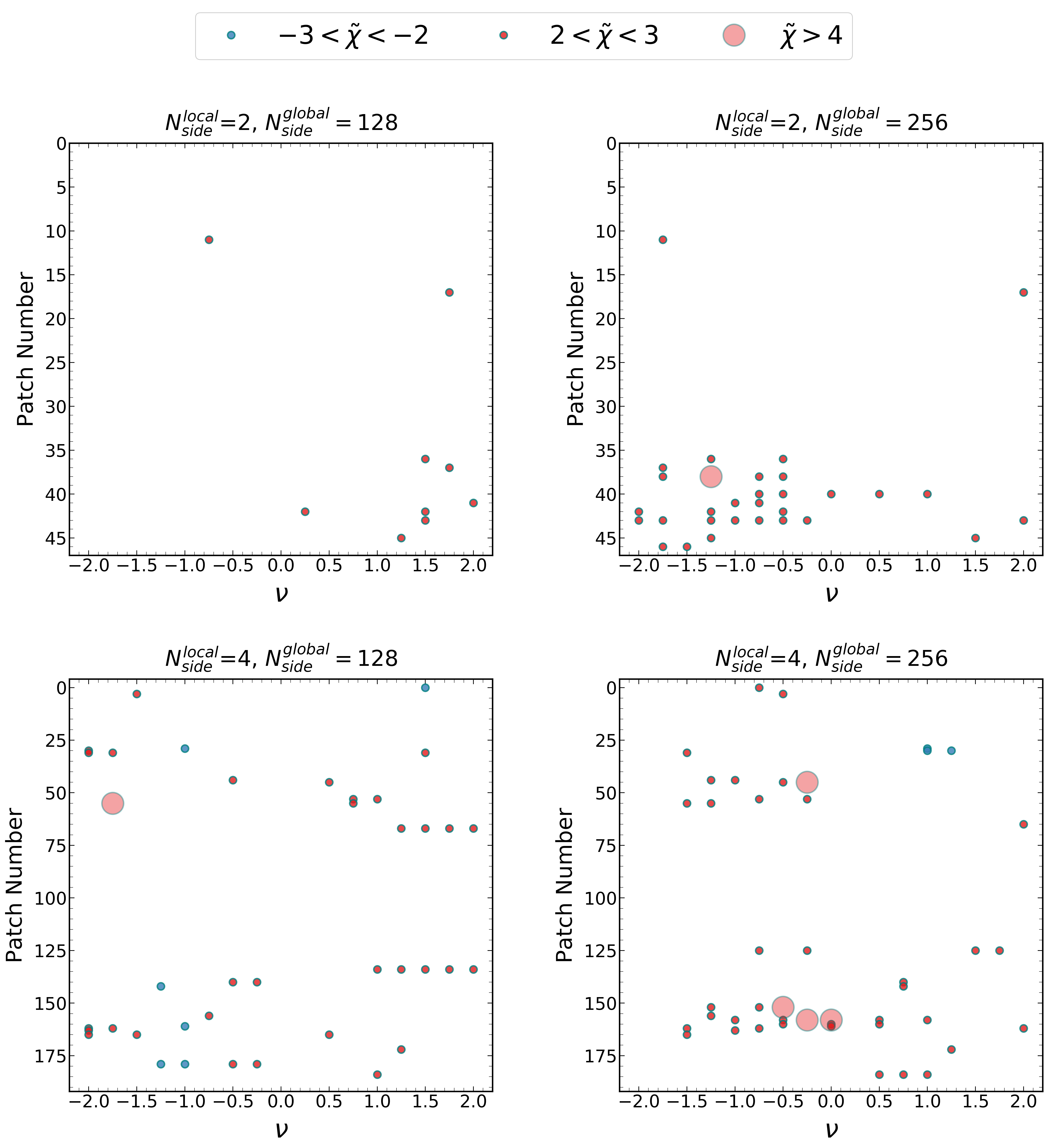}  
\caption{$\widetilde\chi$ values for outlier patches versus threshold.} 
\label{fig:fig5}
\end{figure}


{\em Degree of  anomaly}:  We define a new statistic,  $\widetilde{\chi}(\nu)$, to measure the degree of statistical significance of $\Delta\alpha$ for every outlier patch. It is defined to have value between 2 and 3 if   $\widetilde\chi^{(2)}>1$ and $\widetilde\chi^{(3)}<1$,   between 3 and 4 if   $\widetilde\chi^{(3)}>1$ and $\widetilde\chi^{(4)}<1$, and so on. 
Figure~\ref{fig:fig5} shows plots of  $\widetilde{\chi}(\nu)$ for all the anomalous patches  on the two dimensional space spanned by $\nu$ and the patch identification numbers. The four panels correspond to the four combinations of $N_{\rm side}^{\rm local}$ and $N_{\rm side}^{\rm global}$. Blue circles represent negative, while pink ones represent positive values of $\widetilde\chi$. Larger sized  circles denote larger values of $|\widetilde\chi|$. There are no points having values $3<|\widetilde\chi|<4$ for any of the patches. 
There are no points in the central regions of the panels because these pixels lie in the masked region of the Galactic plane.

We first discuss our findings for the case of $N_{\rm side}^{\rm local}=2$. From table~\ref{tab:table2} and the top panels of figure~\ref{fig:fig5} we observe  the following points: 
\begin{itemize}
\item The number of outlier patches for $N_{\rm side}^{\rm global}=128$ are 8. All of them have $\widetilde\chi$ value between 2 to 3 at one threshold values each, except patch id 42 which is anomalous at two threshold values. 

\item For $N_{\rm side}^{\rm global}=256$ there are 11 outlier patches. All the 8  patches identified in the case of $N_{\rm side}^{\rm global}=128$ form a subset of these. All the common patches have $\widetilde\chi$ value between 2 to 3. Moreover, most of them are found to be anomalous at multiple threshold values (mostly on  the negative side), as can be seen from a comparison of the top panels of figure~\ref{fig:fig5}. This confirms that the anomalous behaviour of the patches is not erased by variation of resolution, and that the anomaly is more evident in the case of the higher resolution. 
Note that the same value of the threshold will not correspond to the same actual field values for $N_{\rm side}^{\rm global}=128$ and 256, due to the field rescaling by the respective standard deviations. A patch exhibiting anomalous behaviour at some threshold will show  similar behaviour  at a threshold value that is shifted with respect to the former. We can see this, for example, for patch id 11 which is anomalous at only one threshold, namely, $\nu=0.75$, for $N_{\rm side}^{\rm global}=128$, while for $N_{\rm side}^{\rm global}=256$ it is anomalous at $\nu=-1.75$.  Moreover, the  threshold shift need not be the same for all anomalous patches since the fields at the two different resolutions are different due to the inclusion of additional multipoles for the higher resolution. 

\item The 3 additional outlier patches found in the higher resolution case indicate new information regarding   disagreement  of the values of $\kappa$ for multipoles higher than 256 between observed data and FFP10 simulations. Patch ids 40 and 46  have $\widetilde\chi$ values between 2 to 3, while patch id 38 is strongly anomalous with $\widetilde\chi>4$.   
\end{itemize}

 Next, we discuss  $N_{\rm side}^{\rm local}=4$. In this case what we are effectively doing is dividing each patch  of  $N_{\rm side}^{\rm local}=2$ into four equal parts and treating each part as a new patch. There is no new information compared to $N_{\rm side}^{\rm local}=2$ in terms of multipoles of $\kappa$. 
 However, the information of anomalous nature of  smaller sized regions can get washed out as components of larger regions since in that case $\alpha$ will capture the mean alignment of a larger number of structures.    
 The analysis for  $N_{\rm side}^{\rm local}=4$ can isolate such smaller anomalous regions.
 From table~\ref{tab:table2} and the bottom panels of figure~\ref{fig:fig5} we observe  the following points: 
\begin{itemize}
\item The number of outlier patches for $N_{\rm side}^{\rm global}=128$ are 22, of which 5 have negative $\widetilde\chi$. All the outliers have $\widetilde\chi$ values between 2 and 3, while  55 is strongly anomalous with $\widetilde\chi>4$. 

\item For $N_{\rm side}^{\rm global}=256$ we get 23 outliers, out of which 2 have negative $\widetilde\chi$ values. All the outliers have $\widetilde\chi$ values between 2 and 3, while  45, 152 and 158 are strongly anomalous with $\widetilde\chi>4$. 

\item There are 19 common patches. Out of these, we find that patch ids 0, 142 and 161 are anomalous at one threshold each, and the sign of  $\widetilde\chi$ is different for different  $N_{\rm side}^{\rm global}$. For this reason we exclude them from the set of outlier patches. 

 \end{itemize}
 
\begin{figure}[]
	\begin{centering}
	{\includegraphics[height=1.6in,width=2.9in]{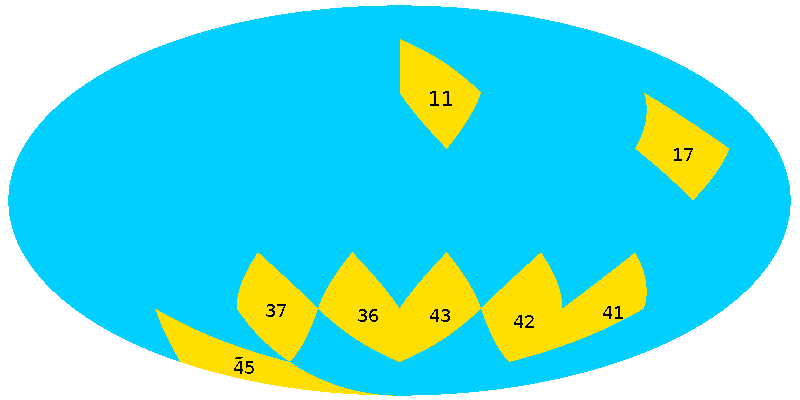}}\hskip .5cm
	{\includegraphics[height=1.6in,width=2.9in]{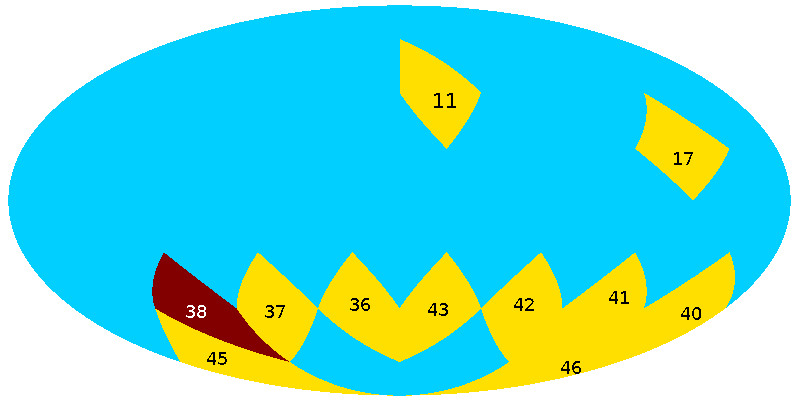}}\\
{\includegraphics[height=1.6in,width=2.9in]{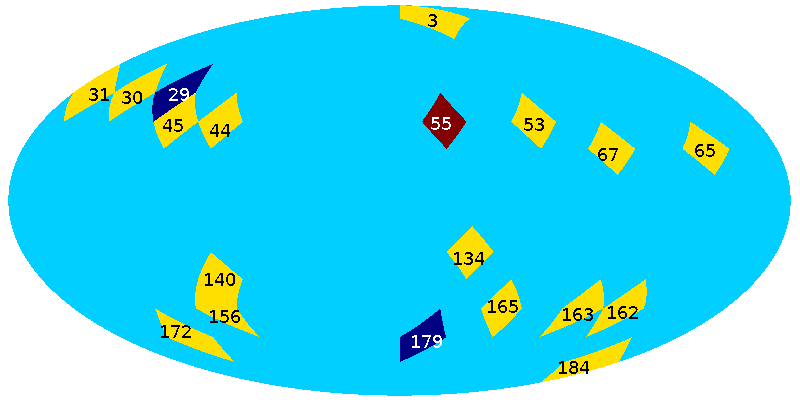}}\hskip .5cm
{\includegraphics[height=1.6in,width=2.9in]{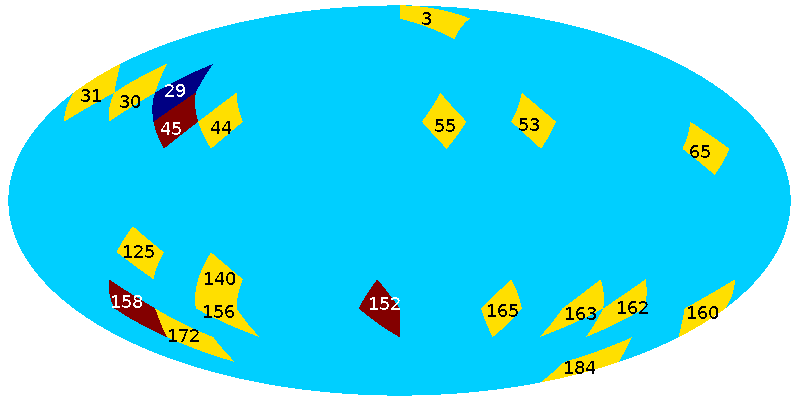}}

\caption{{\em Top}: 
Maps of showing outlier patches for  $N_{\rm side}^{\rm local}=2$ for $N_{\rm side}^{\rm global}=128$ (left) and  $N_{\rm side}^{\rm global}=256$ (right). The colors indicate the corresponding values of  $\widetilde\chi$.  
Dark blue corresponds to $-3 < \widetilde\chi <-2$, yellow corresponds to $2 < \widetilde\chi <3$ and  maroon to $\widetilde\chi > 4$. Patches are numbered in Healpix ring format. 
{\em Bottom}: Same as the top row but for  $N_{\rm side}^{\rm local}=4$.  
All maps are shown using Mollweide projection in Galactic coordinates.}
		\label{fig:fig6}
	\end{centering}
\end{figure}

{\em Sky locations of outlier patches}: Having identified the outlier patches  for each value of $N_{\rm side}^{\rm local}$,
we next examine these patches in further detail by visualising their location in the sky. The top row of figure~\ref{fig:fig6} shows maps of outlier patches, with the different colours indicating the values of $\widetilde\chi$, for  $N_{\rm side}^{\rm local}=2$ for $N_{\rm side}^{\rm global}=128$ (left) and  $N_{\rm side}^{\rm global}=256$ (right). The bottom row shows the same for  $N_{\rm side}^{\rm local}=4$.
Dark blue corresponds to $-3 < \widetilde\chi <-2$, yellow corresponds to $2 < \widetilde\chi <3$ and  maroon to $\widetilde\chi > 4$. Patches ids in Healpix ring format is also indicated. All maps are shown using Molleweide projection in Galactic coordinates.

As seen in figure~\ref{fig:fig6} all the outlier patches for $N_{\rm side}^{\rm local}=2$ and most of them for $N_{\rm side}^{\rm local}=4$ have positive $\widetilde\chi$. As discussed in section~\ref{sec:chi},  positive $\widetilde\chi$ is not an indication of higher level of alignment between the structures. Rather, it indicates  that the observed $\kappa$ field in these patches have higher number of structures in comparison to the simulations. 
Further, we observe a general trend of increase of statistical significance of anomaly for the higher resolution case for both $N_{\rm side}^{\rm local}=2$ and 4. 
For $N_{\rm side}^{\rm local}=2$  the higher resolution case has three additional outlier patches, of which one (patch id 38) is very highly anomalous. For $N_{\rm side}^{\rm local}=4$  the higher resolution case has two additional outlier patches (patch id 152 and 158) which are  highly anomalous, while patch id 45 shows increase of statistical significance of anomaly. 
It is useful to mention here that addition of spatially uncorrelated noise to a given  physical random field having some coherence  length, in general,  results in increase of the number of small scale structures of the combined field.  
Hence, positive values of $\widetilde\chi$ and the general trend for it to be higher for higher resolution correlate with the fact that the SNR of the observed convergence map decreases with increasing multipole.   Therefore, while we cannot conclusively rule out other possibilities, our results strongly suggest that the anomalous behaviour of the outlier patches having positive $\widetilde\chi$ are  caused by inadequate estimation of the instrumental noise.

We find two patches that have negative $\widetilde\chi$  (patch ids 29 and 179 in the bottom left panel of figure\ref{fig:fig6}). These are patches where the structures in the observed $\kappa$ map exhibit relative alignment that is significantly higher (lower value of $\alpha$) than is expected from the simulations. Therefore we conclude that these two patches show departure from SI at higher than 95\% confidence level.   Of these, patch id 29 lies close to the ecliptic north pole, while 179 is located close to the ecliptic plane. 

We can visually see that several of the outlier patches obtained for $N_{\rm side}^{\rm local}=4$ are contained within the outlier patches of $N_{\rm side}^{\rm local}=2$. Thus, as expected, by subdividing the patches of $N_{\rm side}^{\rm local}=2$ into smaller parts we can further isolate or localize the anomalous regions. 
The regions corresponding to patch ids 29, 30, 31, 44 and 25 for $N_{\rm side}^{\rm local}=4$  are conspicuously absent for $N_{\rm side}^{\rm local}=2$. This indicates that the alignment information has gotten washed out when $\alpha$ is calculated over larger patch size. 

Lastly, it is interesting to note that most of the outlier patches are located close to either the ecliptic north or south poles, or the ecliptic plane.  
Patch ids 11 and 17 for $N_{\rm side}^{\rm local}=2$, and correspondingly 53, 55 and 65 for for $N_{\rm side}^{\rm local}=4$, lie close to the masked regions (11 and 55 are close to the north Galactic spur) and are possibly contaminated by residual foreground. 

\section{Conclusion}
\label{sec:sec5}

In this paper we have carried out statistical isotropy test using the observed Planck convergence map by comparing it with  the FFP10 simulations provided by the Planck team. For the test we use the $\alpha$ statistic which measures anisotropy of excursion sets of smooth random fields. We carry  out the test using  the global (masked) sky and for small sky patches. 
In order to focus on multipoles  where the SNR of the observed convergence map is not too small we restrict our analysis to $N_{\rm side}=512$ in the global analysis and 256 in the local analysis. The local analysis is complementary to, but more effective  than, the global one  for detecting presence of anisotropy or any anomalous behaviour of the field in localised sky regions. This is because in carrying out average over a larger set of structures the anomalous information can get washed out. From the global analysis we find that the observed data and simulations show good agreement but hints at the presence of some anomaly. Then using the local patch  analysis we identify several anomalous patches where the observed and simulated data show disagreement of the values of $\alpha$ at statistical significance higher than 95\% CL. 

From the positive sign of the statistic $\widetilde\chi$ we infer that the source of the anomalous behaviour of most of the outlier patches is inaccurate estimation of noise.  
The locations of a majority of the anomalous patches that we have identified are close to either the  ecliptic plane or the ecliptic poles.  
Further we identify two outlier patches,  which exhibit anomalous behaviour originating from departure from SI at higher than 95\% CL. One of these is located near the north ecliptic pole and the other is located close to the ecliptic plane. Though beyond the scope of this paper it will be interesting to examine these patches further and  cross-correlate them with large scale structure surveys. 

Our local patch analysis is similar in spirit to the analysis carried out in Marques et al.~\cite{Marques:2017ejh}, though we use a different methodology. Their method uses the variance, and hence direct field values, computed from sky patches, while $\alpha$ is constructed using the first derivative of the field and can be directly interpreted as the anisotropy of a curve. Therefore, the anomalous patches identified by our analysis need not be the same as those identified using the variance. Comparison of figure~\ref{fig:fig6} with figs.~9 and 10 of \cite{Marques:2017ejh} shows that there are some patches common, and both our analysis and theirs indicate that anomalous regions are more likely near the ecliptic poles and ecliptic plane.   

Our results broadly confirm that the universe is statistically isotropic on large scales by using a method that is different from the ones that have previously been used in the literature. We have argued that most of the anomalous regions we have identified can are well explained by inaccurate noise estimation.  Hence our method and results will be useful for improved understanding of noise. Each of the anomalous regions, and in particular the two regions that exhibit statistically significant deviation from SI, are interesting for further probes and cross-correlation with large scale structure surveys.


\acknowledgments
{We acknowledge  use of the \texttt{NOVA} HPC clusters at the Indian Institute of Astrophysics. Some of the results in this paper have been obtained by using the \texttt{CAMB}~\cite{Lewis:1999b}  and \texttt{HEALPIX}~\cite{Gorski:2005} packages. We have used \texttt{Matplotlib}~\cite{Hunter:2007} library for generating most of the plots in our work. The work of PC is supported by the Science and Engineering Research Board of the Department of Science and Technology, India, under the \texttt{MATRICS} scheme, bearing project reference no \texttt{MTR/2018/000896}. PC would also like to acknowledge the hospitality of the Physics department, Indian Institute for Science Education and Research, Mohali, during the concluding phase of this work.}

\bibliographystyle{JHEP}
\bibliography{main} 

\end{document}